# Comments "On the heat capacity of liquids at high temperatures, S.M. Stishov, Physica A 478 (2017) 205"

I.H. Umirzakov

Institute of Thermophysics, Novosibisk, 630090, Russia

*e-mail: umirzakov@itp.nsc.ru*

**Abstract** It is shown that the isochoric heat capacity of dense gas, fluid and liquid decreases with increasing temperature at arbitrary values of a density for many pair interaction potentials, including bonded potentials; that a decrease of the isochoric heat capacity of the liquid with increasing temperature is related to a decrease of the interaction between the particles with increasing temperature; that a radial distribution function for non-ideal dilute gas, which is independent of density, can describe a temperature dependence of the isochoric heat capacity of liquid argon; that a radial distribution function dependent on the density and temperature describes a temperature dependence of the isochoric heat capacity of liquid and dense fluid; that the Carnahan-Starling equation of state for soft spheres gives a good quantitative description of the isochoric heat capacity of argon; that the fluctuations of the kinetic energy increases with temperature faster than that of the potential energy; and finally, that a liquid state can be considered as a state of dense gas. The explicit expressions to define the Frenkel line on the (temperature, density)–plane are derived. It is shown that the density of liquid argon along the Frenkel line on the (temperature, density)-plane decreases with increasing temperature as $T^{-1/4}$.

## Introduction

As known, the isochoric heat capacity of liquids decreases with increasing temperature [1-10]. A good description of a temperature dependence of the isochoric heat capacity of liquid Ar at high temperatures was achieved [1] earlier using the thermodynamic relation

$$E = \frac{3}{2}kT + \frac{2\pi}{v}\int_0^\infty U(r)g(r,v,T)r^2 dr, \qquad (1)$$

where $E$ is the internal energy per one particle (atoms or molecules), $T$ is the temperature, $r$ is the distance between centers of the masses of two particles, $v$ is the volume per particle, $g(r,v,T)$ is the radial distribution function, $k$ is the Boltzmann constant, and $U(r)$ is the pair interaction potential between two particles. The repulsive part

$$U_{rep}(r) \equiv 4\varepsilon \cdot r_0^{12}/r^{12} \qquad (2)$$

of the Lennard-Johns 12-6 potential $U_{LG}(r) = 4\varepsilon \cdot (r_0^{12}/r^{12} - r_0^6/r^6)$, where $r_0$ and $\varepsilon$ are the positive parameters: $U_{LG}(r_0) = 0$ and $\varepsilon = \min\{U_{LG}(r), r \geq 0\}$, was used in Eq. 1. Besides, the following two implicit assumptions:

**1.** *the radial distribution function is defined by the relation*

$$g(r,v,T) = \exp[-U(r)/kT]; \qquad (3)$$

**2.** *the potential $U(r)$ in Eq. 3 can be replaced by the potential of the soft spheres $U_{ss}(r)$,*

which is defined by

$$U_{ss}(r) = \infty, \quad r < \sigma(T); \quad U_{ss}(r) = 0, \quad r > \sigma(T), \qquad (4)$$

where $\sigma(T)$ is the diameter of the soft sphere which is defined [11] from

$$U_{rep}(\sigma(T)) = kT, \qquad (5)$$

were used in [1].

The relation

$$g(r,T) = 0, \quad r < \sigma(T); \quad g(r,T) = 1, \quad r > \sigma(T), \qquad (6)$$

obtained from Eqs. 3- 4 and Eq. 1 gives the relation

$$C_v = \frac{3}{2}k + \frac{\pi r_0^3}{6v}\left(\frac{4\varepsilon}{kT}\right)^{1/4} \qquad (7)$$

to define the isochoric heat capacity $C_v$.

We discuss an inconsistency of the above approach and show that such an inconsistency can be avoided, as well as the conclusions of [1] can stay valid. We show that the isochoric heat capacity of dense gas, fluid and liquid decreases with increasing temperature at arbitrary values of a density for many pair interaction potentials, including bonded potentials; that an decrease of the isochoric heat capacity of liquid with increasing temperature is related to a decrease of the interaction between the particles with increasing temperature; that the radial distribution function for the non-ideal dilute gas which is independent of the density can describe the temperature dependence of the isochoric heat capacity of the liquid argon; that the radial distribution function dependent on the density and temperature describes the temperature dependence of the isochoric heat capacity of the liquid; that the Carnahan-Starling equation of state for the soft spheres gives a good quantitative description of the isochoric heat capacity of argon; that the fluctuations of the kinetic energy increases with temperature faster than that of the potential energy; and that the liquid state can be considered as a state of a dense gas. The explicit expressions to define the Frenkel line on the (temperature, density)–plane are derived. It is shown that the density of liquid argon along the Frenkel line on the (temperature, density)-plane decreases with increasing temperature as $T^{-1/4}$.

**Comments**

I. If the diameter of the soft sphere defined from the equation $U_{LJ}(\sigma(T)) = kT$, then

$$\sigma(T) = r_0 2^{1/6}(\sqrt{1+kT/\varepsilon}+1)^{-1/6} \qquad (7a)$$

Eqs. 1,3 and 4 give

$$C_v = \frac{3}{2}k + \frac{\pi r_0^3/v}{3\sqrt{2}}\frac{(\sqrt{1+kT/\varepsilon}+1)^{1/2}}{\sqrt{1+kT/\varepsilon}}. \qquad (7b)$$

It is evident from Fig. 1 that Eq. 7b gives a better description of the isochoric heat capacity of Ar than Eq. 7.

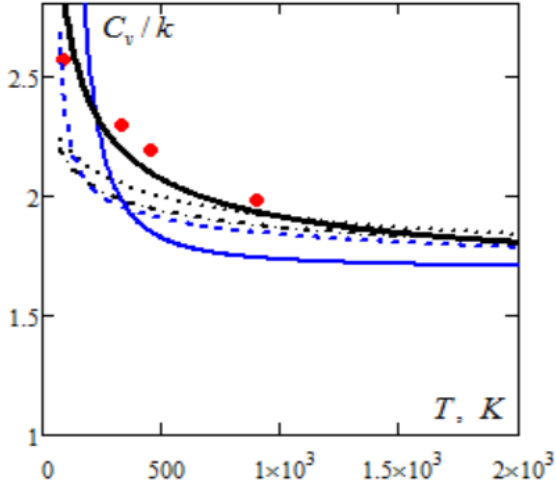

Fig. 1. The temperature dependence of the isochoric heat capacity of liquid Ar at triple point density. Filled red circles correspond to the reference data [12], dashed-dotted black curve – to Eq. 7, black dotted line – to Eq. 7b, solid blue line – to Eq. 8 (Lennard-Jones (12-6) potential, $\varepsilon = 119.3\,K$, $r_0 = 3.405 \cdot 10^{-10}\,m$ [13]), dashed blue line – to Eq. 11, and thick black line – to Eqs. 7a, 15 and 30.

II. There is an inconsistency in the above approach, because two various interaction potentials defined by Eqs. 2 and 4 were used in Eq. 1 to obtain Eq. 7. This inconsistent approach was used in [1] because such a consistent use of the potential of soft spheres (Eq. 2) in Eq. 1 gives the uncertainty $\infty \cdot 0$ at $r = \sigma(T)$.

It is evident that this inconsistency is related to the assumption **2**. Further, we use only assumption **1** to avoid this inconsistency.

Let us consider an arbitrary pair interaction potential $U(r)$. Using the assumption **1**, we have from Eq.1

$$C_v = \frac{3}{2}k + \frac{2\pi}{vkT^2}\int_0^\infty U^2(r)\exp[-U(r)/kT]r^2 dr. \tag{8}$$

Eq. 8 shows that $C_v > 3k/2$ for arbitrary density and temperature. We obtain from Eq. 8

$$\frac{\partial C_v}{\partial T} = -\frac{2\pi k}{vT}\int_0^\infty [2 - U(r)/kT] \cdot [U(r)/kT]^2 \exp[-U(r)/kT]r^2 dr, \tag{9}$$

which shows that $C_v$ decreases with increasing temperature for arbitrary values of the temperature and density if $U(r) \leq 0$ for the arbitrary $r$ and $U_{\min} \equiv \min\{U(r), r \geq 0\} \neq -\infty$ (the non-positive potential) or

$$U(r) = \infty, \quad r < \sigma_0; \quad U(r) = V(r), \quad r \geq \sigma_0, \tag{10}$$

where $V(r)$ is the arbitrary pair interaction potential with non-positive values: $V(r) \leq 0$ at $r \geq \sigma_0$.

One can conclude from Eq. 9 that $C_v$ decreases with increasing temperature at $T \geq U_{\max}$ for a bonded potential, which obeys the condition $U_{\max} \equiv \max\{|U(r)|, r \geq 0\} \neq \infty$. Fig. 2 shows that

$$f(T,n) = \int_0^\infty [2 - U(r,n)/kT] \cdot [U(r,n)/kT]^2 \exp[-U(r,n)/kT]r^2 dr, \quad \text{where}$$

$$U(r,n) = \frac{\varepsilon}{n-6}\left(\left(\frac{r_{00}}{r}\right)^n 6 - \left(\frac{r_{00}}{r}\right)^6 n\right), \tag{9a}$$

is positive in the temperature interval $0 \leq T \leq 500\varepsilon/k$ ($\varepsilon$ is the depth of the potential, $\varepsilon > 0$, $U(r_{00}, n) = -\varepsilon$) for $n = 7, 8, 9, 10, 11, 12, 15, 18, 21, 24, 27, 30, 36$; therefore, according to Eq. 9, the heat capacity decreases with increasing temperature.

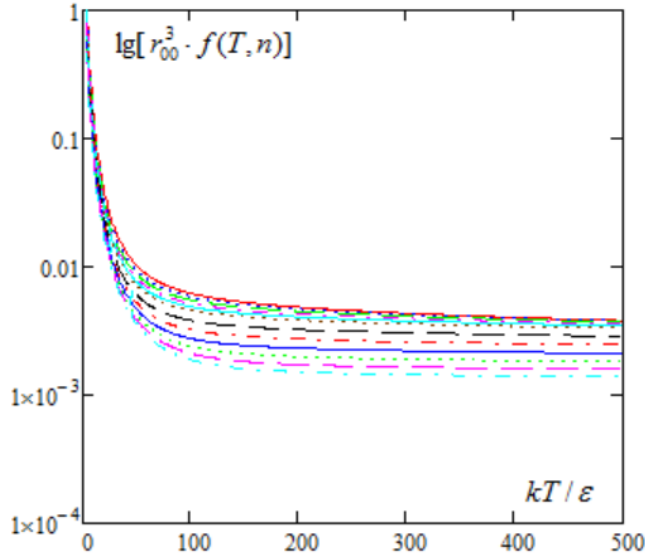

Fig. 2. The dependence of $\lg[r_{00}^3 \cdot f(T,n)]$ on $kT/\varepsilon$ for the pair potential $U(r,n)$, $n = 7, 8, 9, 10, 11, 12, 15, 18, 21, 24, 27, 30, 36$. The more the value of $n$ is, the higher the curve of $\lg[r_{00}^3 \cdot f(T,n)]$ is.

Eq. 8 gives

$$C_v = \frac{3}{2}k + 0.919 \cdot \frac{\pi r_0^3}{6v}\left(\frac{4\varepsilon}{kT}\right)^{1/4} \tag{11}$$

for the repulsive potential defined by Eq. 2. As one can see from the comparison Eq. 7 and 11, the inconsistency mentioned above results in the fact that the contribution of the repulsive forces to the isochoric heat capacity is overestimated by 1.09 times. However, Eq. 7 as well as Eq. 11 qualitatively describe the isochoric heat capacity Ar. Therefore, all conclusions in [1] based on Eq. 7 are valid.

Fig. 1 shows that Eq. 8 for the Lennard-Johns potential qualitatively describes the reference experimental temperature dependence of the isochoric heat capacity of Ar [12]; that Eq. 11 gives a better description of the heat capacity at low temperatures than Eqs. 7 and 7b; and that Eq. 11 for the fully repulsive potential defined by Eq. 2 gives much better description than Eq. 8 for the Lennard-Johns potential.

III. The Frenkel line is defined by the equation $C_v = 2k$ [8, 9, 10, 14]. One can obtain from Eqs. 8 and 11 the corresponding relations

$$n(T) = n_0\left(\frac{kT}{4\varepsilon}\right)^{1/4}, \tag{12}$$

$$\frac{kT(n)}{4\varepsilon} = \frac{n^4}{n_0^4}, \tag{13}$$

$$n(T) = \left(\frac{4\pi}{kT^2}\int_0^\infty U^2(r)\exp[-U(r)/kT]r^2 dr\right)^{-1}, \tag{14}$$

where $n_0 = (0.919\pi r_0^3/3)^{-1}$ for the temperature dependence of the number density $n = 1/v$ along the line. According to [14], the temperature on the Frenkel line of the Lennard-Jones fluid increases very quickly with increasing density. Eq. 13 also predicts a similar behavior of the heat capacity. One can see from Eq. 12 that the density of liquid argon along the Frenkel line on the (temperature, density)-plane decreases with increasing temperature as $T^{-1/4}$.

IV. Eq. 3 is valid for dilute non-ideal gas [15] and Eq. 15 corresponds to considering the second virial coefficient only in the equation $p = \dfrac{kT}{v} - \dfrac{2\pi}{3v^2}\int_0^\infty r\dfrac{dU(r)}{dr}g(r,v,T)r^2 dr$. Therefore, the following question arises: can Eq. 3 take into account the properties of dense gas, fluid and liquid?

Using the exact relation [16]

$$\left(\dfrac{\partial C_v}{\partial v}\right)_T = T\left(\dfrac{\partial^2 p}{\partial T^2}\right)_v \tag{15}$$

and the virial equation of state [17]

$$p = nkT + \sum_{i=2}^{\infty} kTB_i(T)n^i, \tag{16}$$

where $B_i(T)$ is the $i$-th virial coefficient,

$$B_2(T) = 2\pi\int_0^\infty [1 - e^{-U(r)/kT}]r^2 dr, \tag{17}$$

$$B_i(T) = b_i + c_i/T, \quad i \geq 3, \tag{18}$$

$b_i$ and $c_i$ are constants, we obtain Eq. 8 for atomic substance and

$$C_v = C_{v,ig}(T) + \dfrac{2\pi}{vkT^2}\int_0^\infty U^2(r)\exp[-U(r)/kT]r^2 dr \tag{19}$$

for molecular substance. Note that Eqs. 18 are valid at high temperatures [15, 17].

Eq. 18 can be presented as the Van der Waals equation of state [15-18]

$$p = \dfrac{kT}{v-b} - \dfrac{a(T)}{v^2}, \tag{20}$$

if $b_i = b > 0$, $c_i = 0$ for $i \geq 3$, and $a(T) = kT[b - B_2(T)]$.

The virial equation of state Eq. 16 can describe the PVT-properties of dense gas, fluid and liquid [19]. So, we showed that Eq. 3 could take into account the properties of dense gas, fluid and liquid in the isochoric heat capacity.

V. According to Eq. 3, the radial distribution function depends on the temperature but does not depend on the density while it depends on the density and temperature in liquids [15, 20]. We will show in this section that the radial distribution function depending on the density and temperature can describe a temperature dependence of the heat capacity in the liquid.

Let us consider more general case when the third virial coefficient in Eqs. 18 is replaced by the exact relation

$$B_3 = -\frac{8\pi^2}{3} \int_0^\infty x dx \int_0^\infty y dy \int_{|x-y|}^{x+y} f(x)f(y)f(z) z dz, \tag{21}$$

where $f(x) = \exp[-U(x)/kT] - 1$, $x = |\vec{r}_1 - \vec{r}_2|$, $y = |\vec{r}_2 - \vec{r}_3|$, $z = |\vec{r}_3 - \vec{r}_1|$, and $\vec{r}_i$ defines the position of the i-th particle.

We obtain from Eqs. 15-17 and 21

$$C_v = 3k/2 + nk\alpha(T) + n^2 k\beta(T), \tag{22}$$

where

$$\alpha(T) = \frac{2\pi}{k^2 T^2} \int_0^\infty U^2(r) e^{-U(r)/kT} r^2 dr > 0, \tag{23}$$

$$\beta(T) = \frac{4\pi^2}{k^2 T^2} \int_0^\infty x dx \int_0^\infty y dy \int_{|x-y|}^{x+y} [U^2(x)f(y) + 2U(x)U(y)e^{-U(y)/kT}] \cdot e^{-U(x)/kT} f(z) z dz, \tag{24}$$

One can see from Eq. 24 that $\beta(T)$ is a positive function of temperature for a non-positive and bonded potential, and that one defined by Eq. 10. One can conclude from Eqs. 23-24 that $\alpha(T)$ and $\beta(T)$ vanish at high temperatures for above mentioned potentials.

We obtain from Eq. 22 the relation

$$n(T) = \frac{-\alpha(T) + \sqrt{\alpha^2(T) + 2\beta(T)}}{2\beta(T)}, \tag{25}$$

for the temperature dependence of the number density along the Frenkel line.

The rotations of molecule add to the isochoric heat capacity $k$ for linear molecules and $3k/2$ for nonlinear ones. The vibrations of atoms in a molecule also add a corresponding contribution to the heat capacity [16]. Therefore, the Frenkel line for molecular liquids is defined by

$$C_v - C_{v,ig}(T) = k/2, \tag{26}$$

where $C_{v,ig}(T)$ is the isochoric heat capacity per one molecule. One can see from Eqs. 7, 8 and 26 that Eqs. 13, 14 and 25 for the Frenkel line are valid for molecular substances.

We obtain from Eqs. 22-25

$$\frac{\partial C_v}{\partial T} = -2\frac{nk\alpha + n^2 k\beta}{T} + \frac{4\pi^2 n^2}{k^2 T^4} \int_0^\infty x dx \int_0^\infty y dy \int_{|x-y|}^{x+y} [U^2(x) + 2U(x)U(y)]U(y) e^{-[U(x)+U(y)]/kT} f(z) z dz +$$

$$\frac{4\pi^2 n^2}{k^2 T^4} \int_0^\infty x dx \int_0^\infty y dy \int_{|x-y|}^{x+y} [U^2(x)f(y) + 2U(x)U(y)e^{-U(y)/kT}]U(x)e^{-U(x)/kT} f(z) z dz + \tag{27}$$

$$\frac{4\pi^2 n^2}{k^2 T^4} \int_0^\infty x dx \int_0^\infty y dy \int_{|x-y|}^{x+y} [U^2(x)f(y) + 2U(x)U(y)e^{-U(y)/kT}]U(z)e^{-[U(x)+U(z)]/kT} z dz.$$

One can conclude from Eq. 27 that the isochoric heat capacity of liquid decreases with increasing temperature at arbitrary values of the temperature and density for non-positive and bonded potentials, as well as the potential defined by Eq. 10.

We obtain from Eqs. 1a and 15

$$\left(\frac{\partial C_v}{\partial v}\right)_T = -n^2 \frac{2\pi}{3} \int_0^\infty r \frac{dU(r)}{dr} \frac{\partial^2 g(r,v,T)}{\partial T^2} r^2 dr. \tag{28}$$

Eq. 22 gives

$$\left(\frac{\partial C_v}{\partial v}\right)_T = -n^2 k\alpha(T) - 2n^3 k\beta(T). \tag{29}$$

One can conclude from Eqs. 28-29 that the radial distribution function is a linear function of density. So, we have shown that the radial distribution function dependent on density and temperature can describe a temperature dependence of the heat capacity in liquid.

VI. As one can see from Fig. 1, the heat capacity which is defined from Eq. 15 and the Carnahan-Starling equation of state [20,21]

$$p = nkT \frac{1 + \eta + \eta^2 - \eta^3}{(1-\eta)^3}, \tag{30}$$

where $\eta = \pi\sigma^3(T)n/6$ and the diameter $\sigma(T)$ of soft spheres is defined from Eq. 7a, gives a better quantitative description of the heat capacity of Ar than Eqs. 7, 7b, 8 and 11.

We can conclude from the above consideration that the values $C_V$ greater than $2k$ do not correspond to solid-like states of liquid, and the transition from $C_V > 2k$ to $C_V < 2k$ across the Frenkel line defined by the equation $C_V = 2k$ [6] is not a transition from solid-like states to gas-like ones. The above conclusions are in accordance with those in [1].

The Van der Waals, virial and Carnahan-Starling equations of state describe pressure of the ideal, non-ideal and dense gas, fluid and liquid [15-20]. Therefore, taking into account the above consideration we can conclude that a liquid and dense fluid states can be considered as a state of dense gas.

VII. According to [8] $C_V = (\Delta E)^2 / NkT$, where $(\Delta E)^2$ is the mean of the fluctuations of the square of the total energy $E = E_k + V$ of atoms or molecules. In a classical (non-quantum) case the total kinetic $E_k$ and total potential energy of interaction $V$ between atoms are independent, therefore

$$C_V = (\Delta E_k)^2 / NkT + (\Delta V)^2 / NkT = 3kT/2 + (\Delta V)^2 / NkT, \tag{31}$$

where $(\Delta E_k)^2 = 3NkT/2$ and $(\Delta V)^2$ are the mean of the fluctuations of the squares of $E_k$ and $V$, respectively. We can conclude from Eq. 18 that the inequality $C_V < 3k$ valid for liquid and supercritical fluid states [1, 6, 7] means that the fluctuations of the kinetic energy are greater than that of the potential energy. Eq. 31 gives $(\partial C_V / \partial T)_V = 2/3 \cdot (\partial [(\Delta V)^2 / (\Delta E_k)^2] / \partial T)_V$, therefore, the inequality $(\partial C_V / \partial T)_V < 0$ valid for liquid and supercritical states [1, 6, 7] means that the fluctuations of the kinetic energy increases with temperature faster than that of the potential energy.

**Conclusion**

It is shown that the isochoric heat capacity of dense gas, fluid and liquid decreases with increasing temperature at arbitrary values of a density for many pair interaction potentials, including bonded potentials; that a decrease of the isochoric heat capacity of the liquid with increasing temperature is related to a decrease of the interaction between the particles with increasing temperature; that a radial distribution function for non-ideal dilute gas, which is independent of density, can describe a temperature dependence of the isochoric heat capacity of liquid argon; that a radial distribution function dependent on the density and temperature describes a temperature dependence of the isochoric heat capacity of liquid and dense fluid; that the Carnahan-Starling equation of state for soft spheres gives a good quantitative description of the isochoric heat capacity of argon; that the fluctuations of the kinetic energy increases with temperature faster than that of the potential energy; and that a liquid state can be considered as a state of dense gas. The explicit expressions to define the Frenkel line on the (temperature, density)–plane are derived. It is shown that the density of liquid argon along the Frenkel line on the (temperature, density)-plane decreases with increasing temperature as $T^{-1/4}$.

**Highlights**

- The decrease of the isochoric heat capacity of the liquid with increasing the temperature is related to the decrease of the interaction between the particles consisting of liquid with increasing temperature.
- The radial distribution function of the non-ideal gas can describe quantitatively the temperature dependence of the isochoric heat capacity of the liquid argon.
- The Carnahan-Starling equation of state for the soft spheres gives good qualitative description of the isochoric heat capacity of liquid argon.
- The density of liquid argon along the Frenkel line on the (temperature, density)-plane decreases with increasing temperature as $T^{-1/4}$.
- The liquid state can be considered as a state of a dense gas.